\documentstyle[aps,preprint,eqsecnum]{revtex}
\tightenlines
\newcommand{\be}{\begin{equation}}
\newcommand{\ee}{\end{equation}}
\newcommand{\bea}{\begin{eqnarray}}
\newcommand{\eea}{\end{eqnarray}}
\newcommand{\ben}{\begin{eqnarray*}}
\newcommand{\een}{\end{eqnarray*}}
\begin{document}
\title{First Passage Time Problem for Biased Continuous-time Random Walks}
\author{Govindan Rangarajan \thanks{Also associated with the
Jawaharlal Nehru Centre for Advanced Scientific Research,
Bangalore, India; e-mail address: rangaraj@math.iisc.ernet.in}}
\address{Department of Mathematics and Centre for Theoretical Studies\\
Indian Institute of Science \\ Bangalore 560 012, India}
\author{Mingzhou Ding\thanks{E-mail address: ding@walt.ccs.fau.edu}}
\address{Center for Complex Systems and Brain Sciences \\
Florida Atlantic University \\ Boca Raton, FL 33431}
\maketitle

\begin{abstract}

We study the first passage time (FPT) problem for biased continuous
time random walks. Using the recently formulated framework of
fractional Fokker-Planck equations, we obtain the Laplace transform of the FPT
density function when the bias is constant. When the bias depends
linearly on the position, the full FPT density function is
derived in terms of Hermite polynomials and generalized Mittag-Leffler
functions.

\end{abstract}
\newpage

\section{Introduction}

Continuous time random walk (CTRW) \cite{shlesinger} is an
important model for anomalous diffusion \cite{list} in which the
mean square displacement scales with time as $<X^2(t)> \sim
t^{\gamma}$ with $0<\gamma<2$. In applications of CTRW, one is
often interested in its first passage time (FPT) density function
\cite{risken}. The theory of first passage time problems has a
long history
\cite{bachelier,schrodinger,smoluchowski,kramers,darling,montroll,landauer,weiss}
with applications in physics \cite{gardiner}, biology
\cite{tuckwell}, and engineering \cite{cai}.

In an earlier communication \cite{rd}, we had
described the exact solution for the FPT density function
for CTRW (in the diffusion limit) with zero bias.
In this paper, we consider the first passage time problem for
CTRW with nonzero bias (both constant and nonconstant). We start
with a brief description of the CTRW and write down the
fractional Fokker-Planck
equation that describes it following Metzler et. al. \cite{metzler2}.
Next we obtain the Laplace transform of FPT density function for
a constant bias CTRW. Finally we obtain the full FPT density
function when the bias depends linearly on position in terms of Hermite polynomials
and generalized Mittag-Leffler functions \cite{erdelyi}.

\section{CTRW with nonzero bias}

Consider a one dimensional continuous time random walk
on a discrete space lattice. Denoting the probability
of being at site $j$ at time $t$ by $p_j(t)$ and restricting
ourselves to nearest neighbor jumps, the CTRW can be
described by the following generalized master equation \cite{metzler2}:
\be
p_j(t) = \int_0^t dt' \left[ A_{j-1} p_{j-1}(t') + B_{j+1}
p_{j+1}(t') \right] \psi(t-t') + \delta_{x,0} \Phi(t). \ee Here
$\psi(t)$ is the waiting time distribution, $\Phi(t)= 1-\int_o^t
dt' \psi(t')$ is the survival probability, $A_{j-1}$ ($B_{j+1}$)
is the probability to jump from site $j-1$ ($j+1$) to site $j$. We
consider a biased random walk where the probabilities $A_j$ and
$B_j$ are not equal and could depend on the position of the random
walker. However, they fulfill the local condition $A_j+B_j = 1$.
The waiting time distribution is taken to be a Levy distribution
\cite{taqqu} with a power law tail: $\psi(t) \sim
(t/\tau)^{-1-\gamma}$ for large $t$ where $0 < \gamma < 1$ and
$\tau$ has dimensions of time. The biased CTRW described above is
subdiffusive i.e., its mean squared displacement varies with time
$t$ as $t^{\gamma}$ where $0 < \gamma < 1$.

Following Metzler et. al. \cite{metzler2}, we can now transition to
continuous space and introduce the Taylor series expansion:
\be
A_{j-1}p_{j-1}(t) \sim A(x) p(x,t) - \Delta x
\frac{\partial A(x) p(x,t)}{\partial x} + \frac{(\Delta x)^2}{2}
\frac{\partial^2 A(x) p(x,t)}{\partial x^2},
\ee
with a similar expansion for $B_{j+1} p_{j+1}(t)$. We substitute the
above expansions in the generalized master equation and Laplace transform
it in time. Further, we take
the generalized diffusion limit: $\Delta x, \ \tau \to 0$ such that
\bea
F(x) & = & \lim_{\Delta x, \tau \to 0} \frac{\Delta x}{\tau^{\gamma}} [A(x)-B(x)], \\
K & = & \lim_{\Delta x, \tau \to 0} \frac{(\Delta x)^2}{2 \tau^{\gamma}}.
\eea
Here $F(x)$ is the force acting on the particle during every jump and
$K$ is the generalized diffusion constant. Due to the presence of the
force $F(x)$, the biased CTRW is also known as ``CTRW with (non)constant
force'' \cite{metzler2}. The Laplace transformed generalized master equation can be
inverted (see Metzler et. al. \cite{metzler2} for
details) to finally give the following fractional
Fokker-Planck equation (FFPE) for the
probability density function $W(x,t)$ of the
biased CTRW in the generalized diffusion limit:
\be \label{FFPE3}
W(x,t)-W(x,0) =\   _0 D_t^{-\gamma} \left[ K \frac{\partial^2}{\partial x^2} W(x,t)
- \frac{\partial}{\partial x} F(x) W(x,t) \right],
\ \ \ 0 < \gamma < 1.
\ee
Here we have incorporated the initial condition $W(x,0)$ and the Riemann-Liouville
fractional integral operator $_0 D_t^{-\gamma}$ acting on a function $g(t)$
is defined as \cite{oldham,miller}
\begin{equation}
_0 D_t^{-\gamma} g(t) = \frac{1}{\Gamma(\gamma)} \int_0^t \ dt'
\ (t-t')^{\gamma-1}g(t'), \ \ \ \gamma > 0.
\end{equation}
Here $\Gamma(z)$ is the usual gamma function \cite{gradshteyn}. We
call the diffusive process obtained from the CTRW as ``Levy type
of anomalous diffusion'' due to the waiting time distribution
characteristics \cite{rd}.

We now formulate the first passage time problem for the FFPE
given in Eq. (\ref{FFPE3}). This problem can be recast
as a boundary value problem with absorbing boundaries \cite{risken}.
In our case,
to obtain the FPT density function, we first need to solve
Eq. (\ref{FFPE3}) with the following boundary and initial
conditions:
\be \label{bc2}
W(0,t)=0,\ \ \ W(\infty,t) = 0,\ \ \ W(x,0) =\delta(x-a),
\ee
where $x=a$ is the starting point of the CTRW, containing the
initial concentration of the distribution. Once
we solve for $W(x,t)$, the first passage time density $f(t)$
is given by \cite{risken}
\begin{equation}
f(t) = - \frac{d}{dt} \int_0^{\infty}\ dx \ W(x,t).
\label{fptdef}\end{equation}

\section{First Passage Time for Constant Bias}

In this section, we study the first passage time problem for
Levy type anomalous subdiffusion with constant bias $-\nu$
(that is, $F(x)=-\nu$).
The Laplace transform of the FPT density is obtained.

For constant bias, the FFPE to be solved is given by [cf. Eq. (\ref{FFPE3})]
\be \label{FFPE3b}
W(x,t)-W(x,0) = \ _0 D_t^{-\gamma} \left[ \nu \frac{\partial}{\partial x} W(x,t)
+ K \frac{\partial^2}{\partial x^2} W(x,t) \right],
\ \ \ 0 < \gamma < 1.
\ee
with the boundary and initial conditions given in Eq. (\ref{bc2}).
We solve the FFPE using the method of separation of variables \cite{morse}.
Let $W(x,t)=X(x) T(t)$. Substituting in Eq. (\ref{FFPE3b}) we obtain
\be
X(x)T(t) - X(x) = \left[\ _0D_t^{-\gamma} T(t) \right] \left[ \nu X'(x)
+ K X''(x) \right],
\ee
where the primes denote the derivatives with respect to $x$.
Separating out the variables and introducing the separation constant $\lambda$
we get
\be
K X''(x) + \nu X'(x) =  -\lambda X(x),
\label{xeq}\ee
and
\be
T(t)-1  =  -\lambda \ _0D_t^{-\gamma} T(t).
\label{teq}\ee

First we solve Eq. (\ref{teq}).  Taking its Laplace transform we obtain
\be
T(s)-\frac{1}{s} = \frac{\lambda}{s^{\gamma}} T(s).
\ee
Here we have used the fact that the Laplace transform of
$ _0D_t^{-\gamma} T(t)$ is $T(s)/s^{\gamma}$. Solving for $T(s)$ we get
\be
T(s) = \frac{1}{s-\lambda s^{1-\gamma}}.
\ee
Taking the inverse Laplace transform \cite{erdelyi} we finally obtain
\be
T(t) = E_{\gamma}\left[ -\lambda t^{\gamma} \right],
\ee
where $E_{\gamma}(z)$ is the Mittag-Leffler function \cite{erdelyi}
with the following power series expansion:
\be \label{mldef}
E_{\gamma}(z) = \sum_{n=0}^{\infty} \frac{z^n}{\Gamma(1+\gamma n)}.
\ee
Note that $E_{\gamma}(z)$ reduces to the regular exponential function
when $\gamma=1$.

Next consider Eq. (\ref{xeq}). The solution of this equation satisfying
the boundary conditions is given by
\be
X(x) = \exp[-\nu (x-a)/2 K]
\frac{\sin[x \sqrt{\lambda/K-\nu^2/4K^2}]}{2 \sqrt{K \lambda - \nu^2/4}},
\ \ \ \lambda \ge \nu^2/4K.
\ee
Thus we have a continuous spectrum for $\lambda$. Combining the
solutions for $X(x)$ and $T(t)$, $W(x,t)$ is
given by the following integral over $\lambda$:
\be
W(x,t) = \frac{2}{\pi} \int_{\nu^2/4K}^{\infty} d \lambda
A (\lambda) \exp[-\nu (x-a)/2 K]
\frac{\sin[x \sqrt{\lambda/K-\nu^2/4K^2}]}{2 \sqrt{K \lambda - \nu^2/4}}
E_{\gamma}\left[ -\lambda t^{\gamma} \right].
\ee
The coefficient $A(\lambda)$ is fixed by the initial condition
($W(x,0) = \delta (x-a)$) and we get
\bea
W(x,t) & = & \frac{2}{\pi} \int_{\nu^2/4K}^{\infty}  d \lambda
\exp[-\nu (x-a)/2 K] \sin[a \sqrt{\lambda/K-\nu^2/4K^2}] \nonumber \\
& &
\frac{\sin[x \sqrt{\lambda/K-\nu^2/4K^2}]}{2 \sqrt{K \lambda - \nu^2/4}}
 E_{\gamma}\left[ -\lambda t^{\gamma} \right].
\eea

Letting $\lambda' = \sqrt{\lambda/K-\nu^2/4K^2}$ we obtain
\be
W(x,t) = \frac{2}{\pi} \int_0^{\infty} d \lambda'
\sin \lambda' a \sin \lambda' x \exp[-\nu (x-a)/2 K]
E_{\gamma}\left[ -(K \lambda^{\prime 2} + \nu^2/4K) t^{\gamma} \right].
\ee
Using standard trigonometric identities and dropping
the primes, the above equation can be rewritten as
\bea
W(x,t) & = & \frac{1}{\pi} \int_0^{\infty} d \lambda
\cos \lambda (x-a) \exp[-\nu (x-a)/2 K]
E_{\gamma}\left[ -(K \lambda^2 + \nu^2/4K) t^{\gamma} \right] \nonumber \\
& & -\frac{1}{\pi} \int_0^{\infty} d \lambda
\cos \lambda (x+a) \exp[-\nu (x-a)/2 K]
E_{\gamma}\left[ -(K \lambda^2 + \nu^2/4K) t^{\gamma} \right].
\eea
Taking the Laplace transform with respect to time we get
\bea
q(x,s) & = & \frac{1}{\pi} \int_0^{\infty} d \lambda
\cos \lambda (x-a) \exp[-\nu (x-a)/2 K]
\frac{1}{s+\nu^2 s^{1-\gamma}/4K + K \lambda^2 s^{1-\gamma}} \nonumber \\
& & -\frac{1}{\pi} \int_0^{\infty} d \lambda
\cos \lambda (x+a) \exp[-\nu (x-a)/2 K]
\frac{1}{s+\nu^2 s^{1-\gamma}/4K + K \lambda^2 s^{1-\gamma}} ,
\eea
where $q(x,s)$ is the Laplace transform of $W(x,t)$.
Here we have used the fact that the Laplace transform of
$E_{\gamma}(-B t^{\gamma})$ is $1/(s+B s^{1-\gamma})$ \cite{erdelyi}.

We can now perform the integration over $\lambda$ by using the
following result \cite{erdelyi2}
\be
\int_0^{\infty} \frac{ \cos \lambda x }{\alpha^2 + \lambda^2} =
\frac{\pi}{2 \alpha} e^{-\alpha |x|}.
\ee
Thus we obtain
\bea \label{ltp}
q(x,s) & = & \frac{1}{2 \sqrt{K}} \exp[-\nu (x-a)/2 K]
\frac{s^{\gamma-1}}{\sqrt{s^{\gamma}+\nu^2/4K}} \nonumber \\
& &
\left\{ \exp[-\sqrt{s^{\gamma}+\nu^2/4K}\ |x-a|/\sqrt{K} ] -
\exp[-\sqrt{s^{\gamma}+\nu^2/4K}\ (x+a)/\sqrt{K} ] \right \}.
\eea

To obtain the Laplace transform $F(s)$ of the FPT density function $f(t)$,
we take the Laplace transform of Eq. (\ref{fptdef}) to get
\begin{equation}
F(s) = -s \int_0^{\infty} dx \, q(x,s) + \int_0^{\infty} dx \, W(x,0).
\end{equation}
Here we have used the fact that Laplace transform of $dW(x,t)/dt$ is
given by \cite{erdelyi2} $s q(x,s) - W(x,0)$. Since $W(x,0) = \delta(x-a)$
[cf. Eq. (\ref{bc2})], we obtain
\begin{equation} \label{fslt}
F(s) = 1-s \int_0^{\infty} dx \, q(x,s).
\end{equation}
Substituting for $q(x,s)$ from Eq. (\ref{ltp}), we get
\bea
F(s) & = & 1- \frac{1}{2 \sqrt{K}}
\frac{s^{\gamma}}{\sqrt{s^{\gamma}+\nu^2/4K}}
\int_0^{\infty} d x \ \exp[-\nu (x-a)/2 K] \nonumber \\
& & \left\{
\exp[-\sqrt{s^{\gamma}+\nu^2/4K}\ |x-a|/\sqrt{K} ] -
\exp[-\sqrt{s^{\gamma}+\nu^2/4K}\ (x+a)/\sqrt{K} ] \right \}.
\eea
After considerable manipulation, we obtain the final
result upon evaluating the integrals to be
\be
F(s) = \exp \left[a \left( \nu - \sqrt{\nu^2+4 K s^{\gamma}} \right)/2K \right],
\ \ \ 0 < \gamma \le 1.
\ee

We now prove that $F(s)$ is the Laplace transform of a valid probability
distribution by showing that $F(0)=1$ and $F(s)$ is completely monotone \cite{feller}:
\be
(-1)^n \frac{d^n F(s)}{ds^n} \ge 0, \ \ \ \forall n,\ s > 0.
\ee
It is easily verified that $F(0)=1$. We next prove that $F(s)$ is completely monotone
for $0 < \gamma \le 1$.
$F(s)$ can be written as the product $F_1(s) F_2(s)$ where $F_1(s) =
\exp(a \nu /2K)$ and $F_2(s)=\exp(-\rho(s))$ with
\be
\rho(s) = \sqrt{\nu^2+4 K s^{\gamma}}/2K.
\ee
From Feller \cite{feller}, $F_2(s)$ is completely monotone if
$\rho(s)$ is a positive function (for $s > 0$) with a completely monotone
derivative. It is obvious that $\rho(s)$ is a positive function.
The derivative $\rho'(s)$ is given by
\be
\rho'(s) = \frac{\gamma s^{\gamma-1}}{\sqrt{\nu^2+4 K s^{\gamma}}}.
\ee
Now $\rho'(s)$ can be written as the product $\rho_1'(s) \rho_2'(s)$
where
\be
\rho_1'(s) = \gamma s^{\gamma-1},\ \ \ \rho_2'(s) = 1/\sqrt{\nu^2+4 K s^{\gamma}}.
\ee
The first factor $\rho_1'(s)$ is completely monotone since
\be
(-1)^n \frac{d \rho_1'(s)}{ds^n} = (-1)^n \gamma (\gamma-1) \cdots (\gamma-n) s^{\gamma-n-1} \ge 0,
\ee
for $0 < \gamma < 1$. Similarly, $\rho_2'(s)$ is also completely monotone.
Now the product of two completely monotone functions is also completely
monotone \cite{feller}. Hence $\rho'(s)$ is completely monotone. This implies
that $F_2(s)$ is completely monotone. Trivially, $F_1(s)$ is also completely
monotone. Therefore the product $F(s)=F_1(s) F_2(s)$ is completely monotone
for $0 < \gamma < 1$. Consequently, $F(s)$ is the Laplace transform of a
valid probability distribution.

From the above expressions for $F(s)$ and its derivatives, one can calculate
the moments of the FPT distribution \cite{feller}. We find that all the
moments diverge for $0 < \gamma < 1$.

The above solution for $F(s)$ can be extended to include $\gamma=1$ (ordinary
Brownian motion). Specifically, for $\gamma=1$, the inverse
Laplace transform of $F(s)$ can be carried out to give
\begin{equation}
f(t) = \frac{a}{\sqrt{4 \pi K t^3}} \exp \left[ -
\frac{(a-\nu t)^2}{4 K t} \right], \ \ \ a >0, \ \ t > 0.
\end{equation}
This is nothing but the FPT density function for a
Brownian motion with drift $\nu$ \cite{grimmet}.
This can be seen as follows. For $\gamma=1$, the FFPE
in Eq. (\ref{FFPE3b}) reduces to:
\be
W(x,t)-W(x,0) = \ _0 D_t^{-1} \left[ \nu \frac{\partial}{\partial x} W(x,t)
+ K \frac{\partial^2}{\partial x^2} W(x,t) \right].
\ee
Differentiating both sides with respect to $t$ we get
\be
\frac{\partial}{\partial t}W(x,t) = \nu \frac{\partial}{\partial
x} W(x,t) + K \frac{\partial^2}{\partial x^2} W(x,t). \ee This is
the usual Fokker-Planck equation for ordinary Brownian motion with
drift $\nu$ thus explaining the observed FPT density for
$\gamma=1$. In other words, for a ordinary Brownian motion, a
constant bias is equivalent to a constant drift.

\section{First Passage Time for Linear Bias}

In this section, we study the first passage time problem for
Levy type anomalous subdiffusion with bias $-\omega x$ which
depends linearly on the position $x$ (that is $F(x)=-\omega x$).
An expression for the FPT density is obtained in terms of Hermite
polynomials and generalized Mittag-Leffler functions \cite{erdelyi}.

In this case, the FFPE to be solved is given by [cf. Eq. (\ref{FFPE3})]
\be \label{FFPE3c}
W(x,t)-W(x,0) = \ _0 D_t^{-\gamma} \left[ \frac{\partial}{\partial x} (\omega x W(x,t))
+ K \frac{\partial^2}{\partial x^2} W(x,t) \right],
\ \ \ 0 < \gamma < 1.
\ee
with
\be
W(0,t) = W(\infty,t) = 0,\ \ \ W(x,0) = \delta(x-a).
\ee
We solve this FFPE again using the method of separation of variables \cite{morse}.
Let $W(x,t)=X(x) T(t)$. Substituting in Eq. (\ref{FFPE3c}) we obtain
\be
X(x)T(t) - X(x) = \left[\ _0D_t^{-\gamma} T(t) \right] \left[ \omega x X'(x)
+ \omega X(x)+ K X''(x) \right],
\ee
where the primes denote derivatives with respect to $x$.
Separating out the variables and introducing the separation constant $\lambda$
we get
\be
K X''(x) + \omega x X'(x) + \omega X(x) =  -\lambda X(x),
\label{xeq2}\ee
and
\be
T(t)-1  =  -\lambda \ _0D_t^{-\gamma} T(t).
\label{teq2}\ee

The above equations were solved for natural boundary conditions ($W(-\infty,t) =
W(\infty,t)=0$) by Metzler et. al. \cite{metzler2}. We follow a similar approach
here. Let
\be
Y(\tilde{x}) = e^{-\tilde{x}^2/2} X(\tilde{x}),\ \ \ \tilde{x} = \sqrt{\omega/K}\ x.
\ee
Substituting in Eq. (\ref{xeq2}) and assuming $\omega \neq 0$ we get
\be
Y''( \tilde{x}) - \tilde{x} Y'( \tilde{x}) + \frac{\lambda}{\omega} Y( \tilde{x}) = 0.
\ee
This is nothing but the differential equation satisfied by Hermite polynomials
$H_n( \tilde{x}/\sqrt{2})$. In our case we have the boundary conditions
$Y(0) = Y(\infty)=0$. But it can be easily shown that any function satisfying
these boundary conditions can be represented in terms of the odd degree Hermite
polynomials $H_{2n+1}( \tilde{x}/\sqrt{2})$. Further, these polynomials satisfy
the following orthogonality condition which follows in a straightforward fashion
from the usual orthogonality condition \cite{gradshteyn} satisfied over the
domain $(-\infty, \infty)$:
\be \label{ortho}
\int_0^{\infty} dx H_{2n+1}(x) H_{2m+1}(x) e^{-x^2} = \sqrt{\pi} 2^{2n}
(2n+1)! \delta_{nm}.
\ee

Thus the equation for $Y( \tilde{x})$ becomes an eigenvalue equation with
the eigenvalue
\be \label{eigenv}
\lambda_{2n+1} = (2n+1) \omega, \ \ \ n=0,1, \ldots,
\ee
and the eigenfunction
\be
Y_{2n+1}( \tilde{x}/\sqrt{2}) = \frac{1}{2^n \pi^{1/4} \sqrt{(2n+1)!}}
H_{2n+1} (\tilde{x}/\sqrt{2}).
\ee
In terms of the original variables we have
\be
X_{2n+1}(x) = \frac{1}{2^n \pi^{1/4} \sqrt{(2n+1)!}}
e^{-\omega x^2/2K} H_{2n+1} (\sqrt{\omega/2K} x).
\ee

We have already solved the equation for $T(t)$ in the previous
section. The only difference here is that $\lambda$ takes on
discrete values given by Eq. (\ref{eigenv}):
\be
T_{2n+1}(t) = E_{\gamma}[-(2n+1)\omega t^{\gamma}],
\ee
where $E_{\gamma}(z)$ is the Mittag-Leffler function defined
earlier. The general solution for $W(x,t)$ can now be written
down:
\be
W(x,t) = \sum_{n=0}^{\infty} A_{2n+1} \frac{1}{2^n \pi^{1/4} \sqrt{(2n+1)!}}
e^{-\omega x^2/2K} H_{2n+1} (\sqrt{\omega/2K} x) E_{\gamma}[-(2n+1)\omega t^{\gamma}].
\ee
The coefficients $A_{2n+1}$ are determined by imposing the initial
condition $W(x,0) = \delta(x-a)$. Using the orthogonality condition
for $H_{2n+1}(x)$ given in Eq. (\ref{ortho}), we can easily determine
$A_{2n+1}$ to be
\be
A_{2n+1} = \sqrt{\omega/2K} \frac{1}{2^n \pi^{1/4} \sqrt{(2n+1)!}}
H_{2n+1} (\sqrt{\omega/2K} a).
\ee
Substituting this in the expression for $W(x,t)$ we get
\bea
W(x,t) & = & \sqrt{\omega/2 \pi K} \sum_{n=0}^{\infty} \frac{1}{4^n (2n+1)!}
H_{2n+1} (\sqrt{\omega/2K} a) e^{-\omega x^2/2K} \nonumber \\
& & H_{2n+1} (\sqrt{\omega/2K} x) E_{\gamma}[-(2n+1)\omega t^{\gamma}].
\eea

To obtain the FPT distribution, we substitute the above expression for
$W(x,t)$ in Eq. (\ref{fptdef}). Using the following result \cite{gradshteyn}
\be
\int_0^{\infty} dx e^{-\omega x^2/2K} H_{2n+1} (\sqrt{\omega/2K} x) =
\sqrt{2K/\omega} (-1)^n \frac{4^n}{\sqrt{\pi}} \Gamma (n+1/2),
\ee
we get
\be
f(t) = - \frac{1}{\pi} \sum_{n=0}^{\infty} \frac{ (-1)^n \Gamma (n+1/2)}{\Gamma (2n+2)}
H_{2n+1} (\sqrt{\omega/2K} a) \frac{d}{dt} E_{\gamma}[-(2n+1)\omega t^{\gamma}].
\ee
From the series expansion of Mittag-Leffler function Eq. (\ref{mldef}) we have
\be
\frac{d}{dt} E_{\gamma}[-(2n+1)\omega t^{\gamma}] =
\sum_{k=0}^{\infty}
\frac{\gamma k [-(2n+1) \omega t^{\gamma}]^{k-1} [-(2n+1) \omega t^{\gamma-1}]}{\Gamma(1+\gamma k)}.
\ee
Using the facts that the summand vanishes for $k=0$ and $\Gamma(1+x)=x\Gamma(x)$ we have
\be
\frac{d}{dt} E_{\gamma}[-(2n+1)\omega t^{\gamma}] =
\sum_{k=1}^{\infty}
\frac{\gamma k [-(2n+1) \omega t^{\gamma}]^{k-1} [-(2n+1) \omega t^{\gamma-1}]}{\gamma k \Gamma(\gamma k)}.
\ee
Letting $k'=k-1$ and simplifying we get
\be
\frac{d}{dt} E_{\gamma}[-(2n+1)\omega t^{\gamma}] = -(2n+1) \omega t^{\gamma-1}
\sum_{k'=0}^{\infty}
\frac{[-(2n+1) \omega t^{\gamma}]^{k'}}{ \Gamma(\gamma k' + \gamma)}.
\ee
But the infinite sum is the series expansion of the generalized Mittag-Leffler
function $E_{\gamma,\gamma}[ -(2n+1) \omega t^{\gamma}]$ \cite{erdelyi}.
Therefore we have
\be
\frac{d}{dt} E_{\gamma}[-(2n+1)\omega t^{\gamma}] = -(2n+1) \omega t^{\gamma-1}
E_{\gamma,\gamma}[-(2n+1) \omega t^{\gamma}].
\ee
Substituting this in the expression for $f(t)$ we finally get
\be
f(t) = \frac{\omega}{\sqrt{\pi} t^{1-\gamma}}
\sum_{n=0}^{\infty} \frac{ (-1)^n }{4^n n!}
H_{2n+1} (\sqrt{\omega/2K} a) E_{\gamma,\gamma}[-(2n+1)\omega t^{\gamma}].
\ee
Here we have also used the following doubling rule \cite{gradshteyn} for gamma functions
to simplify the expression for $f(t)$:
\be
\Gamma(2n+2) = (2n+1)\Gamma(2n+1) = (2n+1) 4^n \Gamma(n+1/2) \Gamma(n+1)/\sqrt{\pi}.
\ee

\section{Conclusions}

In this paper we studied the first passage time (FPT) problem
for a biased continuous time random walk (CTRW). Solution to
this problem was obtained using the recently formulated
fractional Fokker-Planck equation which describes the CTRW
in the (generalized) diffusion limit. Expressions for the
FPT density function was obtained in two cases -- constant
bias and bias that depends linearly on the position.

\newpage
\section*{Acknowledgements}

This work was supported by US ONR Grant N00014-99-1-0062. GR
thanks Center for Complex Systems and Brain Sciences, Florida
Atlantic University, where this work was performed, for
hospitality.

\end{document}